\documentclass[10pt,a4paper]{article}
\usepackage{epsfig,ifdraft,pdfpages}
\usepackage{epsfig,graphicx,amsfonts,amsmath}

\usepackage{latexsym}
\usepackage{amsmath}

\newlength{\dinwidth}
\newlength{\dinmargin}
\def\eq#1{{Eq.~(\ref{#1})}}
\newcommand{\Le}{\left(}
\newcommand{\Ra}{\right)}

\newcommand{\beq}{\begin{equation}}
\newcommand{\eeq}{\end{equation}}
\newcommand{\beqar}{\begin{eqnarray}}
\newcommand{\eeqar}{\end{eqnarray}}
\newcommand{\D}{\partial}

\newcommand{\ep}{\varepsilon}

\def\1{\partial_1}
\def\2{\partial_2}
\def\+{\partial_+}
\def\-{\partial_-}
\def\ip{\partial_i}

\def\eq#1{{Eq.~(\ref{#1})}}

\usepackage{color}

\date{}

\begin{document}

\title {{~}\\
{\Large \bf  NNLO classical solution for Lipatov's effective action for reggeized gluons }\\}
\author{
{~}\\
{\large
S.~Bondarenko,
S.~Pozdnyakov
}\\[7mm]
{\it\normalsize   Physics Department, Ariel University, Ariel, Israel}\\
}

\maketitle

{\bf Abstract}

We consider the formalism of small-x effective action for reggeized gluons, \cite{Gribov,LipatovEff,BFKL}, and,
following to the approach developed in \cite{Bond2,Bond3}, calculate the classical gluon field to NNLO
precision with fermion loops included.
It is  demonstrated, that the  the self-consistency of the equations of motion in each perturbatie order in the approach  is equivalent to the
transversality conditions applied to the solutions of the equations in the lower orders, that allows to construct the solutions with the help of some recursive scheme.
Applications of the obtained results are also discussed.

\section{Introduction}
$\,\,\,\,\,\,\,\,\,\,$In the framework of small-x Lipatov's effective action, \cite{LipatovEff}, the calculation of the amplitudes of high-energy scattering can be done by two different methods.
The first one, which we can call as diagrammatic, see \cite{BKP, Cherednikov, BKP1, Hameren}, is based on the use of effective vertices of reggeons-gluons interactions
obtained from the action and subsequent construction of the amplitudes with the use of the vertices. Another method, proposed as well in \cite{LipatovEff} and developed
in  \cite{Bond2,Bond3}, is based on the formalism of the effective action. In this case, the field theory is constructed  on the base of the fluctuations around the classical solutions of motion.
Because of the dependence of the classical gluon fields on the reggeon fields,
the result obtained is revealed as the Regge Field Theory (RFT), see \cite{Bond3, OldRFT, NewRFT1}, where the new RFT action is appearing as a functional of the reggeon fields only. In this form the action can be served as a generating functional of the reggeon-reggeon interactions and can be used also for the calculations of the reggeon loops arising in the RFT.
This reformulation of the effective action for reggeized gluons as RFT can be considered as
some generalization of Gribov's Regge calculus, \cite{Gribov}, for the case of QCD degrees of freedom.
The validity of the effective action approach was confirmed in different calculations, see \cite{EffAct,Nefedov:2017qzc}.

 Therefore, the use of the approach in the form considered in \cite{Bond2} requires the knowledge of the classical solutions of the equations of motion for the gluon fields.
In the perturbative scheme, used in \cite{Bond3}, each term of the classical solution represents some bare vertex of reggeon-reggeon interactions, which will affects on the  calculations of the
gluons loops corrections as well. In the calculation of the NLO\footnote{ In the RFT approach the LO contribution is represented by a bare action of the reggeized gluons, whereas the NLO
contribution is given by RFT action with one gluon loops included, i.e. by the action with contribution of the terms quadratic on fluctuations in the Lipatov's effective action included.}
propagator of reggeized gluons the only LO precision for the classical solutions is required, see \cite{Bond2}. Nevertheless, the calculation of the different vertices of reggeon-reggeon interactions with NLO and NNLO precision will require the NLO and NNLO precision for the classical gluon fields, this problem is considered in the present paper. We note also, that the classical solution with
NNLO precision consists  with fermion loops contribution as well, see\cite{Fadin}.

The paper is organizing as following. In the Section 2 we remind the main ideas of the
formalism of the Lipatov's effective action, main results of \cite{Bond2} and
discuss the self-consistency of the equations of motion and it's relations to the transversality condition for a certain combination of fields.
In Sections 3,4 and 5 the calculations of the classical solutions for longitudinal and  transverse components of the gluon field are represented.
Section 6 is dedicated to the inclusion of the quark loop contribution in the approach. The last section is the Conclusion of the paper.

\section{Effective action for reggeized gluons with color field source}

$\,\,\,\,\,\,\,\,\,\,$The Lipatov effective action, see \cite{LipatovEff, OurWithZubkov}, is a non-linear gauge invariant action which describes clusters of particles in correspondence with their rapidities $y=\frac{1}{2} ln  \Big( \frac{p_+}{p_-} \Big)$ whereas the interaction between the clusters with essentially different rapidities is realized by the
reggeon fields exchange. The action has the following form
\begin{equation}
 L\,=\,-\,\frac{1}{4} G_{\nu \mu}^a  G^{\nu \mu}_a\,-\,v_{+}J^{+}\,-\,v_{-}J^{-} \,+\, L_{quark}\,,
 \label{L1}
\end{equation}
where the quark part of the Lagrangian is given by
\beq
L_{quark}\,=\, \sum_{c=1}^{n_f}  \bar{\psi}_c ( i \gamma^{\nu} \partial_{\nu} \,-\, m \,+\, g \gamma^{\nu} T^a v_{a \nu})\psi_c \,.
\eeq\,
Under variation on the gluon fields these currents reproduce the Lipatov's induced currents
\beq
\delta (v_{\pm}J^{\pm})\,=\,\delta (v_{\pm}) j^{\pm}\,,
\eeq
where
\beq\label{Intr1}
j^{\pm}_a = -\ tr (f_a j^{\pm})=  \frac{1}{N} tr (f_a O(v_\pm) f_b O^T(v_\pm)) (\partial_i^2 A_\mp).
\eeq
There are additional kinematical constraints for the reggeon fields
\begin{equation}
\partial_- A_{+} \,=\, \partial_+ A_{-} \,=\, 0 \,,
\label{kin}
\end{equation}
corresponding to the strong-ordering Sudakov components in the multi-Regge kinematics, see \cite{LipatovEff} and also \cite{OurWithZubkov}.

 The equations of motion in light-cone gauge ($v_-^a=v^{+ \ a}=0$) provided by \eq{L1} Lagrangian (without the quark term, see Section 6), have the following form:
\begin{equation}
(D_\mu G^{\mu \nu})_a=\partial_\mu G^{\mu \nu}_a+gf_{abc} v_\mu^b G^{c\  \mu \nu} = j_a^+ \delta^{\nu +} + j^-_a \delta^{\nu -}.
\label{mot}
\end{equation}
We solve these equations perturbatively, order by order, with the help of the usual perturbative expansion of the gluon fields:
\beq\label{Ch1_1}
v_+^a = \sum_{k=0}^{+\infty} g^k v_{+k}^a \,,\,\,\,\, v^{ia} = \sum_{k=0}^{+\infty} g^k v_{k}^{ia}.
\eeq
Here, among of four equations for three fields, one from the equations provides us with a transversality condition for a certain combination of fields. If this condition is satisfied,
then the classical solutions for the fields exist and they have a very simple structure.
Indeed, the same \eq{mot} we can rewrite in an another form. Due the covariant derivative presence, we can separate the terms in the equations and rewrite the equations as some
recurrence relations between the different terms of the \eq{Ch1_1} series.
Namely,
the fields of $k$ perturbative order will stay in the left-hand side of the equations and the terms which consists with the fields of perturbative order less than $k$ but  multiplied on
the corresponding order of coupling constant $g$ and denoted as $\bar{j}_{k-1}^+$, $\bar{j}_{k-1}^-$ and $ \bar{j}_{k-1}^i$
will stay on the right hand side of equations. Therefore, we obtain:
\beq \label{j+}
-\- [\ip v^i_k + \- v_{+k} ] = \bar{j}_{k-1}^+ \,,
\eeq
\beq \label{jj}
\Box v^j_k -\partial^j [\ip v^i_k + \- v_{+k} ] = \bar{j}_{k-1}^j \,,
\eeq
\beq \label{j-}
\Box v_{+k} -\partial_+ [\ip v^i_k + \- v_{+k} ] = \bar{j}_{k-1}^- \,.
\eeq
Taking derivatives of the equations in correspondence to the indexes of the r.h.s of them and
summing up the l.h.s. of the obtained expressions we obtain that the self-consistency of the solutions of the \eq{j+}-\eq{j-} is equivalent to
the condition of the transversality being imposed on the r.h.s of the sum:
\begin{equation}
 \partial_\mu \bar{j}_{k-1}^\mu = 0,
 \label{trans}
\end{equation}
that means that the classical solutions of the lower perturbative  orders and \eq{Intr1} induced current substituted in $\bar{j}_{k-1}^\mu$ functions must have the form which satisfies the
\eq{trans} condition.
This conditions, therefore, provide a useful way for the constructions and check of the classical solutions of the equations of motion.
Indeed, thus we have:
\begin{equation}
 v^j_k = \Box^{-1} \Big[\bar{j}_{k-1}^j -  \partial^j \-^{-1} \bar{j}_{k-1}^+ \Big],
 \label{vjk}
\end{equation}
\begin{equation}
 v_{+k} = \Box^{-1} \Big[\bar{j}_{k-1}^- -  \partial_+ \-^{-1} \bar{j}_{k-1}^+ \Big].
 \label{v+k}
\end{equation}
Therefore, there are
the LO and NLO expressions for the
classical fields obtained in \cite{Bond2}:
 \begin{equation}
  v_{+0  a}= A_{+  a},
  \label{v+0}
 \end{equation}
\begin{equation}
  v_{i0  a}= \rho_i^b (x_{\perp},x^-) U_{ab},
  \label{vi0}
 \end{equation}
\begin{equation}
  v_{+1  a}=-\frac{2}{g} \Box^{-1} \Big( (\+ \partial^i U_{ab} ) \rho_i^b  \Big),
  \label{v+1}
 \end{equation}
\begin{equation}
  v_{i1  a}=-\Box^{-1} \Big[ \partial^j F_{ji  a}\,+\, \frac{1}{g} \ip \Big( ( \partial^j U_{ab})   \rho_j^b \Big)  \,-\, \ip \-^{-1} j^{+}_{a1} \Big],
  \label{vi1}
 \end{equation}
 where
\begin{equation}
 \rho^i_a=-\frac{1}{N} \-^{-1} (\partial^i A_{-  a}),
 \label{rho}
\end{equation}
\begin{equation}
 U^{ab}(v_+) =  tr \Big[ f^a \Big( Pe^{g \int^{x^+}_{- \infty} dx'^+ v_{+c}(x'^+,x^-,x_{\perp})f^c} \Big)  f^b \Big( Pe^{g \int^{+\infty}_{x^+} dx'^+ v_{+d}(x'^+,x^-,x_{\perp})f^d} \Big) \Big]\,,
\label{U}
 \end{equation}
see Appendix A for details.
To NNLO precision, we write our \eq{Ch1_1} ansatz in the following  form:
\begin{equation}
v_{+}^a=A_{+}^a(x_{\perp},x^+)\,+\,gv_{+1}^{a}(x_{\perp},x^-,x^+)\,+\,g^2v_{+2}^{a}(x_{\perp},x^-,x^+),
 \label{v+}
 \end{equation}
\begin{equation}
 v_{i}^a=v_{i0}^a(x_{\perp},x^-,x^+)\,+\,gv_{i1}^{a}(x_{\perp},x^-,x^+)\,+\,g^2v_{i2}^{a}(x_{\perp},x^-,x^+),
  \label{vi}
 \end{equation}
the calculations of these NNLO terms of the gluon fields are presented in the next Sections.

\section{First equation of motion}

$\,\,\,\,\,\,\,\,\,\,$In this section we consider the equation of motion arising after the variation of the
Lagrangian with respect to  $v_{+a}$ field. We obtain:
\beq\label{Fir1}
\partial_\mu G^{\mu +}_a+gf_{abc} v_\mu^b G^{c  \mu +} = j_a^+
\eeq
or
\beq\label{Fir2}
\partial_i G^{i +}_a + \partial_- G^{- +}_a +gf_{abc} v_i^b G^{c  i +} = j_a^+ \,,
\eeq
where the current in the r.h.s. is given by \eq{Intr1}. Using
\beq\label{Fir3}
G^{\mu +}_a \,=\, \partial^\mu v^+_a\, -\, \partial^+ v^\mu_a \,+\,gf_{abc} v^{\mu  b} v^{+  c}\, =\, -\, \partial^+ v^\mu_a \,,
\eeq
we obtain for \eq{Fir2} the following equation:
\begin{equation}
-\partial_i  \partial^+ v^i_a - \partial_-  \partial^+ v^-_a -gf_{abc} v_i^b  \partial^+ v^{i c} = j_a^+ .
\label{NNLO+1}
\end{equation}
Now we substitute inside of it the perturbative expressions of \eq{v+0}-\eq{vi1} obtaining the following terms.
\begin{enumerate}
\item
The first term in the l.h.s of \eq{NNLO+1}:
\begin{equation} \label{3.1}
-\partial^i  \partial^+ v^{a}_i =
-\partial^i  \partial_- v^{a}_{i0} - g \partial^i  \partial_- v^{a}_{i1} - g^2 \partial^i  \partial_- v^{a}_{i2}.
\end{equation}
With the use of \eq{vi0} expressions we obtain:
\begin{equation}  \label{3.2}
 -\partial^i  \partial_- v_{i0  a}
=  -(\partial^i  \partial_- \rho_i^b) U_{ab} - g( \partial_- \rho_i^b) (\frac{1}{g} \partial^i U_{ab}) - g^2( \partial^i \rho_i^b) (\frac{1}{g^2} \partial_- U_{ab}) - g^2 \rho_i^b (\frac{1}{g^2} \partial^i \partial_- U_{ab})\, .
\end{equation}
The first term in the r.h.s of \eq{3.2} is the same as the r.h.s of \eq{NNLO+1}, they both have the $g^{0}$ order precision as it must be for the this order classical solution.
\item
The second term:
\begin{equation} \label{3.4}
- \partial_-  \partial^+ v^-_a \, =\, -\partial_-  \partial^+ \Big(A_{+ a} + gv_{+1  a}
+g^2v_{+2  a} \Big)\,=\,
-\partial_-  \partial^+ \Big(gv_{+1  a}
+g^2v_{+2  a} \Big)\,
\end{equation}
and consists with the $g$ and $g^{2}$ order terms only.
\item
The third term in \eq{NNLO+1} we write in the following form:
\begin{equation} \label{3.5}
-gf_{abc} v_i^b  \partial^+ v^{i c} \, =\, gj^{+}_{a1}+g^2 L^{+}_{a2}  (A_+,\rho)+ O(g^3)\,,
\end{equation}
where
\begin{equation} \label{3.6}
gj^{+}_{a1}\, =\,-gf_{abc} (U^{bb'} \rho^{i}_{b'})(U^{cc'} (\- \rho_{i  c'}))
\end{equation}
and
\begin{equation}
 g^2 L^{+}_{a2} (A_+,\rho)\, =\, -g^2 f_{abc} (v_{i0}^b \- v_{1}^{ic} + v_{i1}^b \- v_{0}^{ic})\, .
 \label{L+}
\end{equation}
\end{enumerate}

Summing up these three terms and rewriting the equation in the form of \eq{j+} we obtain:
 \beq \label{j+1}
- g^2 \partial_- \Big[ \partial^i   v_{i2 a} + \partial_-  v_{+2a} \Big]
  \, =\, g^2  \Big[ ( \partial^i \rho_i^b) (\frac{1}{g^2} \partial_- U_{ab}) \, -\,  L^{+}_{a2} \, -\,  \Box^{-1} \Big[ \frac{2}{g}  \+ \- j^{+}_{a1} \Big] \Big]
	\eeq
or
\begin{equation}
\Big[ \partial^i   v_{i2 a} + \partial_-  v_{+2  a} \Big]\, =\,  \-^{-1} L^{+}_{a2} \, +\,  \Box^{-1} \Big[ \frac{2}{g}  \+  j^{+}_{a1} \Big] \, -\,  \-^{-1}  (( \partial^i \rho_i^b) (\frac{1}{g^2} \partial_- U_{ab}))\,,
\label{nu+}
\end{equation}
where the r.h.s. of the expression consists with the classical solutions of $k=0$ and $k=1$ orders.

\section{The second and third equations of motion}

$\,\,\,\,\,\,\,\,\,\,$The variation of the action with respect to $v_{ia}$ gives
\begin{equation} \label{4.1}
(D_+ G^{+ i})_a \,+\, (D_- G^{- i})_a \,+\, (D_j G^{j i})_a \,=\,0
\end{equation}
or
\begin{equation}
 \Box v^i_a \, -\, \partial^i \Big[ \partial^j   v_{j  a} \,+\, \partial_-  v_{+  a} \Big]  \,+\, (\partial F)^{i}_{a}\,=\,0 \,,
 \label{nui1}
\end{equation}
where
\beq\label{4.2}
 (\partial F)^{i}_{a}  \, =\,  g f_{abc} \Big(v_+^b \- v^{i \ c} \, +\,  \- ( v^{-  b} v^{i  c} ) \, +\,   \partial_j (v^{j  b} v^{i  c})\,+\,
 v^{j  b} \partial_j  v^{i  c} \, -\,  v_j^{b} \partial^i  v^{j  c}  \, +\,  g f^{cde} v^{j  b}  v^j_d v^{i}_e \Big) \,.
\eeq
The \eq{nui1} can be represented in the form of equation \eq{jj} and subsequently it can be written in the form of \eq{vjk}.
Then, the only unknown functions in  \eq{nui1} are to the $g^2$ order, we denote them as $g^2 C^{i}_a$. Calculation of this contribution is made in appendix B, finally
the result for the classical solution of the transverse gluon fields  to NNLO precision reads as
\beqar \label{4.10}
&& \Box v_{2a}^i - \partial^i \Big[ \-^{-1} L^{+}_{a2}  - \-^{-1}  (( \partial^j \rho_j^b) (\frac{1}{g^2} \partial_- U_{ab}))
 )
 \Big]
 + f_{abc} \Big(v_{1+}^b \- v^{i  c}_0 + v_{0+}^b \- v^{i  c}_1 + \- ( v^{b}_{1+} v^{i  c}_0 + v^{b}_{0+} v^{i  c}_1 ) \nonumber  \\
 &&+\,\partial_j (v^{j \ b}_0 v^{i c}_1 \,+\, v^{j  b}_1 v^{i  c}_0)
 \,+\, v_{0j}^{b} ( \partial^j  v^{i  c}_1 \,-\, \partial^i  v^{j  c}_1)
 \,+\, v_{1j}^{b} ( \partial^j  v^{i  c}_0 \,-\, \partial^i  v^{j  c}_0)
 \,+\, f^{cde} v^{b}_{j0}   v^j_{0d} v^{i}_{0e} \Big)  \,+\, C^i_a  \,=\,0
\eeqar
or
\beqar
 v_{2a}^i &=&  \Box^{-1} \Big[ \-^{-1} \partial^i \Big( L^{+}_{a2}  \,-\, ( \partial^j \rho_j^b) (\frac{1}{g^2} \partial_- U_{ab})
 \Big)
\,-\, f_{abc} \Big( \frac{1}{g} v^{b}_{j0} \Big(\partial^j  v^{i  c}_0 \,-\, \partial^i  v^{j  c}_0  \Big) \,+\, v_{0j}^{b} ( \partial^j  v^{i  c}_1 \,-\, \partial^i  v^{j  c}_1) \nonumber \\
&-&  v^{i  c}_0 \- v_{1+}^b \,+\, 2 A_+^b \- v^{i  c}_1
 \,+\, \partial_j (v^{j  b}_0 v^{i  c}_1 \,-\, v^{i  b}_0 v^{j c}_1 )
 \,+\, f^{cde} v^{b}_{j0}  v^j_{0d} v^{i}_{0e} \Big) \Big]\,.
 \label{vi2}
\eeqar
In turn, the substitution of  \eq{vi2} in \eq{nu+}  provides the classical solution for the longitudinal gluon field:
\beqar \label{4.11}
\partial_-  v_{+2 \ a} &=&
2 \Box^{-1}   \+  \Big[ L^{+}_{a2}  -( \partial^j \rho_j^b) (\frac{1}{g^2} \partial_- U_{ab}))
+ \frac{1}{g}j_{a1}^+ \Big]
+ f_{abc} \Box^{-1} \ip \Big( \frac{1}{g} v^{b}_{j0} \Big(\partial^j  v^{i  c}_0 - \partial^i  v^{j  c}_0  \Big) \nonumber\\
&+& v_{0j}^{b} ( \partial^j  v^{i  c}_1 - \partial^i  v^{j  c}_1)
-   v^{i  c}_0 \- v_{1+}^b \,+\, 2 A_+^b \- v^{i  c}_1 \,+\,  f^{cde} v^{b}_{j0}  v^j_{0d} v^{i}_{0e} \Big) \,.
\eeqar
This expression can be simplified if we are shortening the following  expressions:
\begin{equation} \label{4.12}
f^{ij}\,=\,-f^{ji}\,,\, \, \, \, \, (\ip v^{b}_{j0}) f^{ij}\,=\, \frac{g}{2} \frac{(\ip v^{b}_{j0} \,-\, \partial_j v^{b}_{i0})}{g} f^{ij} \,=\,O(g)
\end{equation}
and
\begin{equation} \label{4.13}
 \frac{1}{g} v^{b}_{j0} \ip \Big(\partial^j  v^{i \ c}_0 - \partial^i  v^{j \ c}_0 +g f^{cde}  v^j_{0d} v^{i}_{0e}  \Big) = - v^{b}_{j0} \partial_i F^{ij}_c \,,
\end{equation}
that gives:
\begin{equation} \label{4.14}
  v_{0j}^{b} \ip ( \partial^j  v^{i  c}_1 \,-\, \partial^i  v^{j  c}_1) \,=\,  v^{b}_{j0} \Box^{-1} \partial_k \partial^k  \partial_i F^{ij}_c \,.
\end{equation}
Then we obtain:
\beqar
 v_{+2  a} &=&
 \Box^{-1} \-^{-1}     \Big[
\frac{2}{g}\+ j_{a1}^+
\,+\, f_{abc}   \Big(-v_{i0}^b \- \Big[ 2 \+  v^{ic}_1 \,-\, \partial^i  v_{1+}^c \,+\, 2 \Box^{-1} \partial_+  \partial_j F^{ji}_c  \Big] \nonumber \\
   &+& 2\- v_{i0}^b \+ v^{ic}_1
\,+\,  (\ip v^{i  c}_0) \- v_{1+}^b \,+\, 2 \partial_i (A_+^b \- v^{i  c}_1) \Big) \Big]\,.
\label{v+2}
\eeqar
Substituting \eq{v+1} and \eq{vi1} in \eq{v+2} and preserving there only $g^2$ order terms in we obtain finally:
\begin{equation}
 v_{+2  a} \,=\,
 \Box^{-1} \-^{-1}     \Big[   \frac{2}{g}\+ j_{a1}^+
\,+ \, f_{abc}   \Big( 2\- v_{i0}^b \+ v^{ic}_1
\,+ \, (\ip v^{i  c}_0) \- v_{1+}^b \,+\, 2 \partial_i (A_+^b \- v^{i  c}_1) \Big) \Big] \,,
 \label{v+2f}
 \end{equation}
where $j_{a1}^+$ is defined by equation \eq{3.6}.

\section{The fourth equation of motion}

$\,\,\,\,\,\,\,\,\,\,$The variation of the action with respect to $v_{-a}$ gives
\begin{equation} \label{5.1}
\partial_\mu G^{\mu -}_a\,+\,gf_{abc} v_\mu^b G^{c  \mu -} \,=\, j^-_a\,,
 \end{equation}
which can be rewritten as
\beq \label{5.2}
\partial_i (\partial^i v^{-}_a - \partial^- v^{i}_a + g f_{abc} v^{i}_b v^{-}_c )+\partial_+ \partial^+ v^-_a
 +gf_{abc} v_i^b (\partial^i v^{- c} - \partial^- v^{i  c} + g f^{cde} v^{i}_d v^{-}_e ) + gf_{abc} v_+^b \partial^+ v^{-  c} = \ip \partial^i A_{+a}
\eeq
or
\begin{equation}
 \Box v_{+a} - \partial_+ \Big[\partial_i  v^{i}_a + \partial_- v_{+a}\Big]
 +gf_{abc} \Big(\partial_i (  v^{ib} v_{+}^c)+ v_i^b (\partial^i v^{c}_+ - \partial_+ v^{i  c} + g f^{cde} v^{i}_d v_{+e} ) +v_+^b \partial_- v^{c}_+ \Big)\,=\, \ip \partial^i A_{+a} \,.
 \label{NNLO-1}
\end{equation}
Now we can verify that the equations of motion \eq{NNLO+1} and \eq{NNLO-1} give us the result \eq{v+2f} for $v_{+2}$ found above.
In the same time this calculation can serve as test of the fulfillment of the condition of the transversality introduced above. The reduction of the equation of motion (\ref{NNLO-1}) to the form of \eq{j-} is presented in the appendix C, we write here the result:
\beqar
 &\,& \Box v_{+2a} - \partial_+ \Big[\partial_i  v^{i}_{2a} + \partial_- v_{+2a}\Big] \nonumber \\
 &=& -\,f_{abc} \Big( -v_{i0}^b  \partial_+ v^{i \ c}_1
 - 2 \partial_i ( A_{+}^b v^{ic}_1 ) + A_{+}^b \Box^{-1} \ip \partial^i  \-^{-1}   j^{+c}_{1}  ) \Big)
\,+\,\Box^{-1} \ip \partial^i  \-^{-1} (\frac{1}{g}  \partial_+   j^+_{a1})  \,.
 \label{NNLO-2}
\eeqar
Then, substituting expression from \eq{nu+} and \eq{L+}, we obtain
\beqar  \label{5.8}
 v_{+2a} &=& \Box^{-1} \-^{-1}   \Big[   \Big[ \frac{1}{g}  \+  j^{+}_{a1} \Big] \,-\, \partial_+  (( \partial^i \rho_i^b) (\frac{1}{g^2} \partial_- U_{ab}))  \nonumber \\
 &+&f_{abc} \Big(
 \- v_{i0}^b \+ v_1^{ic} \,+\,
   2 \partial_i ( A_{+}^b \- v^{ic}_1 ) \,-\, A_{+}^b \Box^{-1} \ip \partial^i  j^{+c}_{1}  ) \Big)
 \Big]\,.
\eeqar
It is easy to show that $\+ j^{+}_{a1}= gf_{abc}f^{cde}A_{+}^b(v_{iod} \- v_{0e}^i)=-gf_{abc} A_{+}^b j_{1}^{+c}$ up to the second order of g. We have:
\begin{equation}  \label{5.9}
\+ j^{+}_{a1}= g(f_{aec}f^{edb}+f_{abe}f^{edc})v_{+}^b(v_{iob} \- v_{0c}^i).
\end{equation}
Using
\begin{equation}  \label{5.10}
 f_{abe}f^{edc} = f_{ade}f^{ebc} - f_{aec}f^{edb},
\end{equation}
we, therefore, obtain:
\begin{equation}  \label{5.11}
\+ j^{+}_{a1}= gf_{abe}f^{edc}v_{+}^b(v_{iod} \- v_{0c}^i)= -gf_{abc} v_{+}^b j_{1}^{+c},
\end{equation}
that finally gives the same as \eq{v+2f} expression.

\section{One-loop quark corrections}

 The full NNLO contribution to the classical solutions of the gluon equations of motion must include the contribution from the fermion loop as well, see \eq{L1}.
Expanding this part of the Lagrangian around $\psi\,=\,\bar{\psi}\,=\,0\,
\footnote{In the approach we separate the classical solutions for the quark fields and fluctuations around these classical solution, in general the non-zero
classical quark fields correspond to the production amplitudes in the approach.}$ classical solutions we have:
\begin{equation}
L_{quark}^{\ep}\,=\,   \sum_{c=1}^{n_f} \bar{\ep}_{\psi  c} ( i \gamma^{\nu} \partial_{\nu} \,-\, m \,+\,  g \gamma^{\nu} T^a v_{a \nu})\ep_{\psi  c} \,=\,  \sum_{c=1}^{n_f} \bar{\ep}_{\psi  c} ( M_0 \,+\, M_1 )\ep_{\psi  c} \,.
\label{Lq}
\end{equation}
The bare quark Green's function of the problem is defined as usual:
\beq \label{6.1}
M_{0x} G^0_{q} (x,y) \,=\, \delta^{(4)} (x-y)
\eeq
and has the following momentum representation:
\beq \label{6.2}
G_q^0(x,y)\,=\, (i\gamma_\nu \partial^\nu_x +m) \Delta_{xy}, \, \, \, \, \, \Delta_{xy}\,=\, \int  \frac{d^4p}{(2\pi)^4} \frac{e^{ip(x-y)}}{p^2-m^2+i0} \,.
\eeq
Below we will use the following properties of the $\Delta$ function:
\beq \label{6.3}
\partial^{\nu}_x \Delta_{xy} =
- \partial^{\nu}_y \Delta_{yx}, \, \, \, \, \, \partial^{\mu}_y \partial^{\nu}_x \Delta_{xy} = \partial^{\nu}_y \partial^{\mu}_x \Delta_{xy} \,.
\eeq
The integration on the fluctuations
\beqar\label{6.4}
 &\,&\int \prod_{b=1}^{n_f} d \bar{\ep}_{\psi \, b} d \ep_{\psi \, b} \ exp \Big[  \sum_{c=1}^{n_f} i \bar{\ep}_{\psi \, c} ( M_0 + M_1 )\ep_{\psi \, c} \Big]
 =  \, \Big( det \Big( M_0 + M_1 \Big) \Big)^{n_f} =  \,  exp \ \Big[n_f tr \ ln \ \Big( M_0 + M_1 \Big) \Big] \,\nonumber \\
 & = & exp \ \Big[ n_f \, tr \ ln \ \Big(  M_0 \Big) \Big] * exp \ \Big[n_f \, tr \ ln \ \Big( 1 +M_0^{-1}  M_1\Big) \Big]
\eeqar
results as the usual additional contribution to action:
\begin{equation}
\Gamma_q\, =-i \, n_f \ tr \Big( ln \ \Big( 1 +M_0^{-1}  M_1\Big)  \Big) \,.
\label{Gq}
\end{equation}
Expanding \eq{Gq} we preserve the only $g^{2}$ order contribution:
\begin{equation} \label{6.5}
 i\,n_f \, \frac{g^2}{4} tr \Big(  G^0_{q} (y,x)  (\gamma^{\nu} v_{a \nu}(x)) G^0_{q} (x,y)  (\gamma^{\mu} v_{\mu}^a(y) )\Big) \,,
\end{equation}
that provides the following quark current to the equation of motion:
\begin{equation}
j_{quark \ a}^{\rho}(z)\,=\,i\, n_f \, \frac{g^2}{2}  tr \Big(  \gamma^{\rho} G^0_{q} (z,y)  \gamma^{\mu} v_{a \ \mu}(y) G^0_{q} (y,z) )\Big) \,.
\label{jq1}
\end{equation}
In order to construct additional parts to the classical gluon fields arising from this current
we have to verify that the condition $\partial_\rho j_{quark \ a}^{\rho}=0 $ is satisfied here.
If it does, these additional parts can be calculated simply as
\begin{equation}
 v^i_{q2a} \,=\, \Box^{-1} \Big[j_{quark \ 2a}^i \,-\,  \partial^i \-^{-1} j_{quark \ 2a}^+ \Big]
 \label{vqi2}
\end{equation}
and
\begin{equation}
 v_{q+2a} \,=\, \Box^{-1} \Big[j_{quark \ 2a}^- \,-\,  \partial_+ \-^{-1} j_{quark \ 2a}^+ \Big]\,,
 \label{vq+2}
\end{equation}
see \eq{vjk}-\eq{v+k}.
The check of $\partial_\rho j_{quark \ a}^{\rho}=0 $ we begin rewriting \eq{jq1} in the following form:
\beq
j_{quark \ a}^{\rho}(z)= -2in_f g^2 \Big(g^{\rho\nu} g^{\mu \sigma} + g^{\rho \sigma} g^{\mu \nu} - g^{\rho \mu } g^{\sigma \nu} \Big) \int d^4y (\partial_{\nu z} \Delta_{zy}) v_{\mu a}(y)
(\partial_{\sigma y} \Delta_{yz})
+ 2in_f g^2 m^2 \int d^4y  \Delta_{zy} v^{\rho}_{a}(y) \Delta_{yz}\,.
\eeq
Then we have
\beqar
 \partial_{\rho z} j_{quark \ a}^{\rho}(z) & = &- 2in_f \, g^2  \int d^4y  \Big(
 (\partial_{\rho z} \partial^\rho_z \Delta_{zy}) v^\nu_{a}(y) (\partial_{\nu y} \Delta_{yz}) +(\partial_{\nu y} \Delta_{yz}) v^\nu_{a}(y) (\partial_{\rho z} \partial^\rho_{ z} \Delta_{zy}) \nonumber \\
&+&  (\partial^\rho_z \Delta_{zy}) v^\nu_{a}(y) (\partial_{\nu z} \partial_{\rho  y} \Delta_{yz}) +(\partial_{\rho z} \partial_{\nu z} \Delta_{zy}) v^\nu_{a}(y) (\partial^\rho_{ y} \Delta_{yz}) \nonumber  \\
&-& \partial_{\rho z} \Big[ (\partial_{\nu z} \Delta_{zy}) v^\rho_{a}(y) (\partial^\nu_{y} \Delta_{yz}) \Big]
 -m^2 \partial_{\rho z} \Big[ \Delta_{zy} v^{\rho}_{a}(y) \Delta_{yz} \Big] \Big)\,.
\eeqar
The second and third lines of this expression are identically equal, and the difference between the first and fourth lines is given by an expression that is also zero:
\beq
\partial_{\rho z} j_{quark \ a}^{\rho}(z)\,=\,-\,2i\,n_f\, g^2 (v^\nu_{a}(z) (\partial_{\nu z} \Delta_{zy})_{y=z}\, +\, (\partial_{\nu z} \Delta_{zy})_{y=z} v^\nu_{a}(z))\,=\,0\,,
\eeq
that completes the check and justifies the \eq{vqi2}-\eq{vq+2} expressions.

 Inserting obtained classical gluon fields solutions in the \eq{L1} action, we will obtain the
action which will  depend only on the reggeon fields. The contribution of the quarks to the reggeon propagator can be calculated similarly to the done in \cite{Bond3} with
$g^2$ order of the contributions preserved.
The components of the field strength tensor with LO precision reads as
\beq\label{RFA2}
G_{+\,-\ 0}^{a}\,=\,0\,,\,\,G_{i\,+\ 0}^{a}\,=\,\D_{i}\,A_{+}^{a}\,,\,\,
G_{i\,-\ 0}^{a}\,=\,-\,\D_{-}\,\textsl{v}_{i0}^{a}\,,\,\,G_{i\,j\ 0}\,=\,0\,,
\eeq
and with quarks contributions included as
\beq\label{RFA2q}
G_{+\,-\ q2}^{a}\,=\,0\,,\,\,G_{i\,+\ q2}^{a}\,=g^2 \,\D_{i}\,v_{q+2}^{a}\,,\,\,
G_{i\,-\ q2}^{a}\,=\,-\,g^2\D_{-}\,\textsl{v}_{qi2}^{a}\,,\,\,G_{i\,j\ q2}\,=\,0\,.
\eeq
It gives for the additional contribution to the Lagrangian of the reggeon fields:
\beq\label{RFA3}
\frac{1}{2}\,G_{\mu\,\nu\ 0}\,G^{\mu\,\nu}_{q2}\,=g^2\,\Le\,\D_{-}\,\textsl{v}_{i0}^{a}\,\Ra\,
\Le\,\D_{i}\,v_{q+2}^{a}\,\Ra\,+g^2\,\Le\,\D_{-}\,v_{qi2}^{a}\,\Ra\,
\Le\,\D_{i}\,A_{+}^{a}\,\Ra\,.
\eeq
Therefore, there is the additions contributions to the effective action to NNLO precision obtained from the inclusion of quarks in the approach:
\beq\label{RFA4}
S_{eff\ q2}=\,-\frac{g^2}{N}\,\int d^{4}x \Le\,\Le\D_{i} v_{q+2}^{a}\Ra\,
\Le \D_{i} A_{-}^{b} \Ra\,+N\,\Le\D_{i} A_{+}^{a}\Ra\,\Le \,\D_{-}\,v_{qi2}^{b} \Ra\,+\,
v_{q+2}^{a} \,\Le \D_{\bot}^{2}\,A^{b}_{-}\Ra\,\Ra\,,
\eeq
which can be written as
\beq\label{RFA5}
S_{eff\ q2}=\,-\frac{g^2}{N}\,\int d^{4}x \Big[\D_{i} \Le\,v_{q+2}^{a} \D_{i} A_{-}^{b} \Ra\,+ \\
N\,\Le \,\D_{-} \Le \,v_{qi2}^{b} \D_{i} A_{+}^{a}\Ra\, - v_{qi2}^{b} \Le \D_{i} \D_{-} A_{+}^{a}\Ra\, \Ra\, \Big] =0,
\eeq
i.e. this additional part of the action does not contribute to the reggeon fields propagator. Nevertheless, there are additional contributions to the propagator arising
from the modified classical solutions. We can calculate them following to the calculations of \cite{Bond3} and determining the additional contribution
to the kernel of interacting of reggeized gluons:
\beq\label{EK3}
\Le\,K^{a\,b}_{x\,y}\,\Ra^{+\,-}_{q1}\,=\,\Le\,\frac{\delta^{2}\,\Gamma_q}{\delta A_{+\,x}^{a}\,\delta A_{-\,y}^{b}}\,\Ra_{A_{+},\,A_{-}\,=\,0}\,.
\eeq
Using \eq{Lq}-\eq{Gq}, the new contribution to this kernel can be written as
\begin{equation} \label{12}
-2i\,n_f\, g^2 \delta^{ab} \int \, d^4 t \, d^4 z \,\Big[ \partial_{i t} \Big( (\partial_t^+ \Delta_{tz})(\partial_z^i \Delta_{zt}) + (\partial_t^i \Delta_{tz})(\partial_z^+ \Delta_{zt}) \Big) \Big] \tilde{G}^{-0}_{t^- y^-}  \delta^2_{t_\perp  y_\perp} \delta^2_{z_\perp  x_\perp} \delta_{z^+  x^+}.
\end{equation}
In the momentum space this expression has the following form:
\begin{equation} \label{13}
-2i\,n_f\, g^2 \delta^{ab} \int \,\frac{d^4 p}{(2 \pi)^3} \,\frac{d^4 q}{(2 \pi)^4}\, d^4 t \, \frac{-ie^{-i(p-q)(t-x)} (p_i-q_i)(p_- q^i + q_- p^i)}{(p^2+i0)(q^2+i0)} \theta (t^- - y^-)  \delta^2_{t_\perp  y_\perp}  \delta_{p_- q_-}\,,
\end{equation}
which
after the integration on $t^+$ and $t_\perp$ reduces to the following integral:
\begin{equation} \label{14}
-2i\,n_f\, g^2 \delta^{ab} \int \, \frac{d^4 p}{(2 \pi)^2}\, \frac{d^4 q}{(2 \pi)^4}\, dt^- \,\theta (t^- - y^-)   \frac{i e^{-i(p-q)(y-x)} (p_i-q_i)(p_- q^i + q_- p^i)}{(p^2+i0)(q^2+i0)}  \delta_{p_- q_-} \delta_{p_+ q_+}\,,
\end{equation}
and which one after the integration on $p_-$ and $p_+$ variables reads as
\begin{equation} \label{15}
-2i\,n_f\, g^2 \delta^{ab} \int \, \frac{d^2 p_\perp}{(2 \pi)^2} \, \frac{d^4 q}{(2 \pi)^4} \, \frac{ie^{-i(p-q)_i(y-x)^i} q_- (p_\perp^2-q_\perp^2)}{(2q_+ q_- -p^2_\perp+i0)(2q_+ q_- -q^2_\perp+i0)}  \int dt^- \theta (t^- - y^-)\,,
\end{equation}
where the integrand of $q_+$ decreases at infinity as $1/q_+^2$ and its poles are located on one side of the real axis that gives
\beq
\Le\,K^{a\,b}_{x\,y}\,\Ra^{+\,-}_{q1}\,=0.
\eeq
That means that the reggeon propagator does not change from taking the quark loop contribution into account, that is well known result, see \cite{Fadin}.

\section{Conclusion}

 In this paper we calculated classical solutions of the equations of motion for gluon field to NNLO precision basing on the
Lipatov's effective action, which can be considered as extension of QCD for the case of high energy interactions, see \cite{OurWithZubkov}.
In the light cone gauge, for the three unknown components of the gluon field, four equations of motion exist. It demonstrated, that
in the perturbative scheme of the solution of the equations, the existing of the solutions is equivalent to the transversality condition \eq{trans} applied to
the current built on found solutions of the lower order. Namely, beginning from the bare effective gluon current in  Lagrangian \eq{L1} and LO solutions
of the equations of motion, the contributions to the higher order of the classical solutions can be constructed with the help of \eq{vjk}-\eq{v+k} if condition \eq{trans}
is satisfied for the constructed current, see also \eq{vqi2}-\eq{v+2} where the fermion loop contribution to the NNLO classical solution is accounted.

 The importance of the found solutions is then that they contribute to the  construction of the QCD based Regge Field Theory (RFT). Considering the reggeized gluons
$A_{+}$ and $A_{-}$ as the main degrees of freedom
at high energy QCD interactions, the found solutions provide the NNLO structure of the effective action of RFT
\beq
\Gamma\,=\,\sum_{n,m\,=\,0}\,\Le\,A_{+}^{a_{1}}\,\cdots\,A_{+}^{a_{n}}\,K^{a_1\,\cdots\,a_{n}}_{b_1\,\cdots\,b_{m}}\,A_{-}^{b_{1}}\,\cdots\,A_{-}^{b_{m}}\,\Ra\,,
\eeq
see calculations in \cite{Bond2, Bond3}. The action of the interacting reggeized gluons in this form allows to calculate the effective vertices of interacting reggeized gluons and
correlation functions of the theory, that is important task from the point of view of unitarization of the amplitudes at high energy interactions. In this set-up, therefore, the NNLO
classical solutions provides some complex bare reggeon-reggeon interactions vertices which can be used further for the calculations of the reggeon loops contribution into the amplitudes
and calculations of the corrections to the simpler bare reggeon-reggeon interactions vertices.

 Finally we conclude, that the classical solution for the gluon field calculated is the first step toward the calculations of the unitarization corrections in the framework of QCD RFT
and we hope that it can be useful also in the calculations performing in the framework of the CGC approach, \cite{Venug, Kovner}, where this solution can provide some unitarization corrections as well.

\newpage
\section*{Appedix A: expressions for $\- U^{ab}$ and $\ip U^{ab}$ functions}
\renewcommand{\theequation}{A.\arabic{equation}}
\setcounter{equation}{0}
$\,\,\,\,\,\,$ For the arbitrary representation of gauge field $v_{+}\,=\,\imath\,T^{a}\,v_{+}^{a}$ with
$D_{+}\,=\,\D_{+}\,-\,g\,v_{+}$, we can consider
the following representation of $O$ and $O^{T}$ operators\footnote{Due the light cone gauge we consider here only $O(x^{+})$ operators.
The construction of the representation of the $O(x^{-})$ operators can be done similarly. } introduced in \cite{Bond2}:
\begin{equation} \label{a1}
O_x \,=\, \delta^{ab}\,+\,g \int d^4 y G^{+aa_1}_{xy} (v_+(y))_{a_1 b} \,=\, 1\,+\,g G^+_{xy} v_{+y}
\end{equation}
and correspondingly
\begin{equation} \label{a2}
O^T_x \,= \,  1\,+\,g v_{+y} G^+_{yx}\,,
\end{equation}
which is redefinition of the operator expansions used in \cite{LipatovEff} in terms of Green's function instead
integral operators.
The Green's function in above equations we understand as Green's function of the $D_{+}$ operator
and express it in the perturbative sense as
\begin{equation} \label{a3}
G^+_{xy}=G^{+0}_{xy}+g G^{+0}_{xz} v_{+z} G^{+}_{zy}
\end{equation}
and
\begin{equation} \label{a4}
G^+_{yx}=G^{+0}_{yx}+g G^{+}_{yz} v_{+z} G^{+0}_{zx}
\end{equation}
with the bare propagators defined as (there is no integration on $x$ variable)
\begin{equation} \label{a5}
\D_{+ x}\,\,G_{x y}^{+\,0}\,=\,\delta_{x\,y}\,,\,\,\,G_{y x}^{+\,0}\,\overleftarrow{\D}_{+ x}\,=\,-\delta_{x\,y}\,.
\end{equation}
Then we have
\begin{equation} \label{a6}
\partial_{kx} G_{xy}^+ = \partial_{kx} G^{+0}_{xy}+g G^{+0}_{xz} \partial_{kz} v_{+z} G^{+}_{zy}
\end{equation}
for $k=-,i\,$, the difference is only $\partial_{-}  v_{+0}=0$, hence $\partial_{-}  v_{0}=O(g)$. Then we obtain
\begin{equation} \label{a7}
\partial_{kx} O_x \,=\, g \partial_{kx} G^+_{xy} v_{+y} \,=\,  -\, g G^{+0}_{xy} (\partial_{ky} v_{+y} \,+ \, g \partial_{kz} (v_{+y} G^{+}_{yz}  v_{+z}))
\,=\, -\, g G^{+0}_{xy} \Big( (\partial_{ky}  v_{+y}) O_y\, +\,  v_{+y} (\partial_{ky} O_y) \Big)\,.
\end{equation}
Thus in NNLO we have
\begin{equation} \label{a8}
\partial_{-x} O_x\, = g^2 G^{+0}_{xy} \Big(\frac{1}{g} \partial_{-y}  v_{+y}\Big)O_y\,, \ \ \ \ \  O_x^T \overleftarrow{\partial}_{- x}\, = -g^2 O_y^T \Big(\frac{1}{g} \partial_{-y}  v_{+y}\Big) G^{+0}_{yx}\,,
\end{equation}
this form is more convenient because $\partial_{-}  v_{+0}=0$. Correspondingly for transverse derivative
\begin{equation} \label{a9}
\partial_{ix} O_x\, =\, g G^{+0}_{xy} \Big(\partial_{iy}  v_{+y}O_y\Big)\,, \ \ \ \ \ O_x^T \overleftarrow{\partial}_{i x}\, =\, -\,g \Big(O_y^T v_{+y} \overleftarrow{\partial}_{iy} \Big) G^{+0}_{yx}\,.
\end{equation}
We are interested in the order of g for the operators  $\partial_{-x} U^{ab}_x$ and $\partial_{ix} U^{ab}_x$.
For $U^{ab}_x=  tr \Big( f^a O_x f^b O^T_x \Big)$, see \cite{Bond2} for details, we can write
\begin{equation} \label{a10}
 \partial_{-x} U^{ab}_x \,=\,  g^2 \ tr \Big( f^a  G^{+0}_{xy} (\frac{1}{g} \partial_{-y} v_{+y})O_y f^b O^T_x \,-\, (\frac{1}{g} \partial_{-y}  v_{+y}) G^{+}_{yx} f^a O_x f^b O_y^T  \Big)\,,
 \end{equation}
\begin{equation} \label{a11}
 \partial_{ix} U^{ab}_x \,=\,  g \ tr \Big( f^a  G^{+0}_{xy} ( \partial_{iy} v_{+y} O_y) f^b O^T_x \,-\,  G^{+0}_{yx} f^a O_x f^b ( O_y^T v_{+y} \overleftarrow{\partial}_{iy}) \Big)\,.
 \end{equation}

\newpage
\section*{Appedix B:  NNLO terms contribution in  \eq{4.1}}
\renewcommand{\theequation}{B.\arabic{equation}}
\setcounter{equation}{0}

$\,\,\,\,\,\,$ In this appendix we present a calculation of the NNLO terms, denoted as  $C^{i}_a$, in \eq{4.1} equation of motion. Subsituting LO and NLO classical solutions into the \eq{4.1} we obtain:
\begin{equation}
 \Box v^i_a - \partial^i \Big[ \partial^j   v_{j \ a} + \partial_-  v_{+ \ a} \Big]  + (\partial F)^{i}_{a}=0,
 \label{nui1d}
\end{equation}
where
\beq\label{d.1}
 (\partial F)^{i}_{a}  =  g f_{abc} \Big(v_+^b \- v^{i \ c} + \- ( v^{- \ b} v^{i \ c} )+  \partial_j (v^{j \ b} v^{i \ c})\,+\,
 v^{j \ b} \partial_j  v^{i \ c} - v_j^{b} \partial^i  v^{j \ c} +g f^{cde} v^{j \ b}  v^j_d v^{i}_e \Big).
\eeq
The expression in the square brackets can be written as:
\begin{equation} \label{4.3}
\Big[ \partial^j   v_{j \ a} + \partial_-  v_{+ \ a} \Big]=\Big[ \partial^j   v_{j0 \ a} + \partial_-  A_{+ \ a} \Big]+ g\Big[ \partial^j   v_{j1 \ a} + \partial_-  v_{+1 \ a} \Big] +g^2
\Big[ \partial^j   v_{j2 \ a} + \partial_-  v_{+2 \ a} \Big]\,.
\end{equation}
Using \eq{nu+} we obtain:
\beq \label{nui2}
-\partial^i \Big[ \partial^j   v_{j \ a} + \partial_-  v_{+ \ a} \Big] =  -\partial^i  \partial_j   v^j_{0 \ a} +
g \partial^i \Big[ \frac{1}{g}(\partial^j U^{ab})\rho_j^{b} -  \-^{-1} j^+_{a1} \Big] \nonumber \\
-g^2 \partial^i \Big[ \-^{-1} L^{+}_{a2}  - \-^{-1}  (( \partial^i \rho_i^b) (\frac{1}{g^2} \partial_- U_{ab}))  \Big].
\eeq
The LO terms are canceled in the equations due
\begin{equation}
 \Box v^i_{0\ a}  - \partial^i  \partial_j   v^j_{0 \ a}=2 \+ \- v^i_{0\ a} + \partial_j \Big( \rho^{i \ b}( \partial^j U_{ab}(A_+))-\rho^{j \ b}( \partial^i U_{ab}(A_+)) \Big)
 \label{nui3}
\end{equation}
with
\begin{equation} \label{4.4}
 \partial^j v^i_{0a} - \partial^i v^j_{0a}=\rho^{i \ b}( \partial^j U_{ab}(A_+))-\rho^{j \ b}( \partial^i U_{ab}(A_+)) \,.
\end{equation}
After the substituting of the NLO classical solutions NLO from \eq{v+1} and \eq{vi1} into the equations, there are the NNLO terms remain
which we  denote as $g^2 C^i_a$:
\beqar
 &&g\Box v^i_{1\ a} +g \partial^i \Big[ \frac{1}{g}(\partial^j U^{ab})\rho_j^{b} - \-^{-1} j^+_{a1} \Big]\nonumber\\
 &+& 2 \+ \- v^i_{0\ a}  + \partial_j \Big( \rho^{i \ b}( \partial^j U_{ab}(A_+))-\rho^{j \ b}( \partial^i U_{ab}(A_+)) \Big)
 \nonumber \\
 &+& g f_{abc} \Big(2A_+^b \- v^{i \ c}_0+  \partial_j (v^{j \ b}_0 v^{i \ c}_0)
 +  v^{b}_{j0} (\partial^j  v^{i \ c}_0 - \partial^i  v^{j \ c}_0) \Big) = g^2 C^i_a \,.
 \label{nui4}
\eeqar
The first line of the expression \eq{nui4} can be simplified by substituting \eq{vi1}
\begin{equation} \label{4.5}
g\Box v^i_{1\ a} +g \partial^i \Big[ \frac{1}{g}(\partial^j U^{ab})\rho_j^{b} - \-^{-1} j^+_{a1} \Big] =- g \partial_j F^{ji}_a.
\end{equation}
With the help of \eq{vi0}  and \eq{U} expressions we can write:
\beqar
\+  U^{ab}(v_+) & = &
 tr \Big[ (f^a f^c - f^c f^a) \Big( Pe^{g \int^{x^+}_{- \infty} dx'^+ v_{+e}(x'^+,x^-,x_{\perp})f^e} \Big)  f^b \Big( Pe^{g \int^{+\infty}_{x^+} dx'^+ v_{+d}(x'^+,x^-,x_{\perp})f^d} \Big)  \Big]  g v_{+c} \nonumber \\
& = & -g  f^{acf} v_{+c} U^{fb}(v_+)
\label{dU}
\eeqar
and
\begin{equation} \label{4.6}
2 \+ \- v^i_{0\ a} =- 2 g f_{abc} \- ( v_+^b v_{0}^{i\ c})\,.
\end{equation}
Then, the \eq{nui4} acquires the following form:
\beqar \label{4.7}
& - & g \partial_j F^{ji}_a
 \,-\, 2 g^2 f_{abc} \- ( v_{+1}^b v_{0}^{i\ c})  \,+\, \partial_j \Big( \rho^{i \ b}( \partial^j U_{ab}(A_+))\,-\,\rho^{j \ b}( \partial^i U_{ab}(A_+)) \Big) \nonumber \\
 & + & g f_{abc} \Big(
   \partial_j (v^{j \ b}_0 v^{i \ c}_0)
 \,+ \,v^{b}_{j0} (\partial^j  v^{i \ c}_0\, -\, \partial^i  v^{j \ c}_0) \Big)
\, = \, g^2 C^i_a\,,
\eeqar
 where
\begin{equation} \label{4.8}
 \partial_j F^{ji}_a =  \partial_j \Big[ \Big( \rho^{i \ b}(\frac{1}{g} \partial^j U_{ab}(A_+))-\rho^{j \ b}(\frac{1}{g} \partial^i U_{ab}(A_+)) \Big) + f_{abc} v^{j \ b}_0 v^{i \ c}_0 \Big],
\end{equation}
and correspondingly
\begin{equation} \label{4.9}
C^{i}_a = f_{abc} \Big[   v^{b}_{j0} \Big( \rho^{i \ b}(\frac{1}{g} \partial^j U_{ab}(A_+))-\rho^{j \ b}(\frac{1}{g} \partial^i U_{ab}(A_+))
\Big) - 2 \- ( v_{+1}^b v_{0}^{i\ c}) \Big].
\end{equation}

\newpage
\section*{Appedix C:  NNLO terms contribution in  \eq{5.1}}
\renewcommand{\theequation}{C.\arabic{equation}}
\setcounter{equation}{0}

$\,\,\,\,\,\,\,$ In this appendix we present the fourth equation (\ref{NNLO-1}) of motion to the form of \eq{j-}.
We write \eq{NNLO-1} by analogy with the calculations from the appendix B in the form
\beqar
&& \Box v_{+2a} - \partial_+ \Big[\partial_i  v^{i}_{2a} + \partial_- v_{+2a}\Big] + \bar{C}^{i}_a \nonumber \\
&& +f_{abc} \Big(\partial_i (  v^{ib}_0 v_{+1}^c)+ v_{i0}^b (\partial^i v^{c}_{+1} - \partial_+ v^{i \ c}_1 +  f^{cde} v^{i}_{0d} A_{+e} )
 + \partial_i (  v^{ib}_1 A_{+}^c)+ v_{i1}^b \partial^i A^{c}_+
  +A_+^b \partial_- v^{c}_{+1} \Big)  =0 \,,
\eeqar \label{5.5}
where
\beq \label{5.3}
g\bar{C}^{i}_a\, =\, \Box v_{+1a}  - \partial_+ \Big[\partial_i  v^{i}_{1a} + \partial_- v_{+1a}\Big] - \frac{1}{g} \partial_+ \Big[\partial_i  v^{i}_{0a} \Big]
+f_{abc} \Big(\partial_i (  v^{ib}_0 A_{+}^c)+ v_{i0}^b (\partial^i A^{c}_+ - \partial_+ v^{i \ c}_0)  \Big) \,.
\eeq
With the help of \eq{vi1} and \eq{v+1} we rewrite the first three terms in \eq{5.3} and obtain:
\beq \label{5.4}
g\bar{C}^{i}_a = -\frac{2}{g} \Big( (\+ \partial^i U_{ab} ) \rho_i^b  \Big) - \Box^{-1} \ip \partial^i \partial_+   \-^{-1} j^+_{a1} - \frac{1}{g} (\partial_i \rho^{ib})  \Big[ \partial_+  U_{ab} \Big]
 +f_{abc} \Big(\partial_i (  v^{ib}_0 A_{+}^c)+ v_{i0}^b (\partial^i A^{c}_+ - \partial_+ v^{i \ c}_0)  \Big) \,.
\eeq
Applying \eq{dU} expression, we see that  the following  terms of $g^2$ order are not canceled in \eq{5.4}:
\begin{equation}
\bar{C}^{i}_a\, =\, f_{abc} \Big(  A^b_+ (\frac{1}{g} \partial^i  U^{cd} ) \rho_{id} +  \partial^i (v^b_{+1} U^{cd} ) \rho_{id} + \partial^i (v^b_{+1} v^{c}_{i0}) - v_{i0}^b (\frac{1}{g} \partial_+ v^{i \ c}_0) \Big) \\
 \,- \,  \Box^{-1} \ip \partial^i  \-^{-1} (\frac{1}{g}  \partial_+   j^+_{a1}) \,.
 \label{g^2}
\end{equation}
 Using \eq{dU} and the antisymmetry property $f^{abc}$, we note that:
\begin{equation}
 f_{abc} \Big(\partial_i (  v^{ib}_0 v_{+1}^c) + \partial^i (v^b_{+1} v^{c}_{i0})\Big)=0 \,,
 \label{N1}
\end{equation}
\begin{equation}  \label{5.6}
 v_{i0}^b f^{cde} v^{i}_{0d} A_{+e} - v_{i0}^b (\frac{1}{g} \partial_+ v^{i \ c}_0) =  O(g) \,,\,\,\,\,
f_{abc} \Big( v_{i0}^b \partial^i v^{c}_{+1} +  \partial^i (v^b_{+1} U^{cd} ) \rho_{id} \Big) = O(g) \,
\end{equation}
and
\beqar  \label{5.7}
 &\,& f_{abc} \Big( \partial_i (  v^{ib}_1 A_{+}^c)+ v_{i1}^b (\partial^i A^{c}_+)
  +A_+^b \partial_- v^{c}_{+1}
 +  A^b_+ (\frac{1}{g} \partial^i  U^{cd} ) \rho_{id} \Big) \nonumber \\
 &=& f_{abc} \Big( - 2 \partial_i ( A_{+}^b v^{ic}_1 )
  +A_+^b \Big[ \partial_- v^{c}_{+1} + \partial^i v^{c}_{i1} \Big]
 +  A^b_+ (\frac{1}{g} \partial^i  U^{cd} ) \rho_{id} \Big)  \nonumber \\
 &=& f_{abc} \Big( - 2 \partial_i ( A_{+}^b v^{ic}_1 ) + A_{+}^b \Box^{-1} \ip \partial^i  \-^{-1}   j^{+c}_{1} \Big) \,.
\eeqar
Thus, with the use of \eq{g^2}-\eq{5.7} we rewrite \eq{NNLO-1} in the following form
\beqar
 &\,& \Box v_{+2a} - \partial_+ \Big[\partial_i  v^{i}_{2a} + \partial_- v_{+2a}\Big] \nonumber \\
 &=& -\,f_{abc} \Big( -v_{i0}^b  \partial_+ v^{i \ c}_1
 - 2 \partial_i ( A_{+}^b v^{ic}_1 ) + A_{+}^b \Box^{-1} \ip \partial^i  \-^{-1}   j^{+c}_{1}  ) \Big)
\,+\,\Box^{-1} \ip \partial^i  \-^{-1} (\frac{1}{g}  \partial_+   j^+_{a1})  \,,
 \label{NNLO-c}
\eeqar
where $j_{a1}^+$ is defined by equation \eq{3.6}. The r.h.s. of the \eq{NNLO-c} consists with the classical solutions of $k=0$ and $k=1$ orders only.

\end{document}